%% file: DeClope_isbra.tex
\documentclass{llncs}

\usepackage{amsfonts,amsmath,amssymb}
\usepackage{graphicx}
\usepackage{listings}
\usepackage{xspace}
\usepackage{url}
\usepackage{fourier}

\newif\ifrevision

\newcommand{\norm}[1]{\left\lVert#1\right\rVert}

\newcommand{\DeCo}{\text{\ttfamily DeCo}}
\newcommand{\DeClone}{\text{\ttfamily DeClone}}

\newcommand{\GLos}{\text{\sffamily GLoss}}
\newcommand{\Spec}{\text{\sffamily Spec}}
\newcommand{\GDup}{\text{\sffamily GDup}}
\newcommand{\Extant}{\text{\sffamily Extant}}
\newcommand{\ADup}{\text{\sffamily ADup}}
\newcommand{\AGain}{\text{\sffamily AGain}}
\newcommand{\ALos}{\text{\sffamily ALoss}}
\newcommand{\ABreak}{\text{\sffamily ABreak}}

\newcommand{\Sig}{\sigma}
\newcommand{\citep}[1]{\cite{#1}}

\begin{document}

\title{Assessing the robustness of parsimonious predictions
  for gene neighborhoods from reconciled phylogenies}

\author{Ashok Rajaraman\inst{1}\and Cedric Chauve\inst{1} \and Yann Ponty\inst{1,2,3}}

\institute{
Department of Mathematics, Simon Fraser University, Burnaby, Canada
\and
Pacific Institute for Mathematical Sciences, CNRS UMI3069, Vancouver, Canada
\and 
CNRS/LIX, Ecole Polytechnique, Palaiseau, France}

\maketitle{}

\begin{abstract} 
The availability of many assembled genomes opens the way to study the
evolution of syntenic character within a phylogenetic context. The
DeCo algorithm, recently introduced by B\'erard \emph{et al.},
computes parsimonious evolutionary scenarios for gene adjacencies,
from pairs of reconciled gene trees.  Following the approach pioneered
by Sturmfels and Pachter, we describe how to modify the DeCo dynamic
programming algorithm to identify classes of cost schemes that
generate similar parsimonious evolutionary scenarios for gene
adjacencies. We also describe how to assess the robustness, again to
changes of the cost scheme, of the presence or absence of specific
ancestral gene adjacencies in parsimonious evolutionary scenarios. We
apply our method to six thousands mammalian gene families, and show
that computing the robustness to changes of cost schemes provides
interesting insights on the DeCo model.
\end{abstract}

\section{Introduction}\label{sec:introduction}
\input{introduction_isbra.tex}

\section{Preliminary: models and problems}\label{sec:preliminaries}
\input{preliminaries_isbra.tex}

\section{Methods}\label{sec:methods}
\input{methods_isbra.tex}

\section{Results}\label{sec:results}
\input{results_isbra.tex}

\section{Discussion and Conclusion}\label{sec:conclusion}\label{sec:discussion}
\input{discussion_isbra.tex}

\section*{Acknowledgement}
The authors wish to express their gratitude towards Anatoliy Tomilov (Ural Federal University) for contributing the open source C++ 11 implementation of the {\ttfamily QuickHull} algorithm used for the convex hull computations within \DeClone.
This research was supported by an NSERC Discovery Grant to C.C.,
    an SFU Michael Stevenson scholarship to A.R.

\bibliographystyle{abbrv}
\bibliography{polytope,deco,paleo}


\end{document}

%% file: introduction_isbra.tex

Reconstructing evolutionary histories of genomic characters along a
given species phylogeny is a long-standing problem in computational
biology. This problem has been studied for several types of genomic
characters (DNA sequences and gene content for example), for which
efficient algorithms exist to compute parsimonious evolutionary
scenarios. Recently, B\'erard \emph{et
al.}~\cite{DBLP:journals/bioinformatics/BerardGBSDT12} extended the
corpus of such results to syntenic characters. They defined a model
for the evolution of gene adjacencies within a species phylogeny,
together with an efficient dynamic programming (DP) algorithm,
called \DeCo{}, to compute parsimonious evolutionary histories that
minimize the total cost of gene adjacencies gain and break, for a
given cost scheme associating a cost to each of these two
events. Reconstructing evolutionary scenarios for syntenic characters
is an important step towards more comprehensive models of genome
evolution, going beyond classical sequence/~content frameworks, as it
implicitly integrates genome
rearrangements~\citep{DBLP:books/daglib/p/ChauveEGST13}. Application
of such methods include the study of genome rearrangement rates and
the reconstruction of ancestral gene order.  Moreover, \DeCo{} is the
only existing tractable model that considers the evolution of gene
adjacencies within a general phylogenetic framework; so far other
tractable models of genome rearrangements accounting for a given
species phylogeny are either limited to single-copy genes and ignore
gene-specific
events~\citep{DBLP:journals/tcbb/BillerFM13,DBLP:journals/bmcbi/TannierZS09},
assume restrictions on the gene duplication events, such as
considering only whole-genome duplication
(see~\cite{DBLP:journals/bmcbi/GagnonBE12} and references there), or
require a dated species phylogeny~\cite{DBLP:journals/jcb/MaRRSZMH08}.

The evolutionary events considered by \DeCo{}, gene adjacency gain and
break caused by genome rearrangement, are rare evolutionary events 
compared to gene-family specific events.  It is then important to
assess the robustness of inferences made by \DeCo{}, whether it is of
a parsimony cost or of an individual feature such as the presence of a
specific ancestral adjacency. We recently explored an approach that
considers the set of all possible evolutionary scenarios under a
Boltzmann probability distribution for a fixed cost scheme~\cite{DBLP:conf/wob/ChauvePZ14}. 
A second approach consists of assessing how robust features of evolutionary scenarios are
to changes in the cost associated to evolutionary events (the cost
scheme). Such approaches have recently been considered
for the gene tree reconciliation problem and have been shown to
significantly improve the results obtained from purely parsimonious
approaches~\citep{DBLP:journals/jcb/BansalAK13,DBLP:journals/bioinformatics/Libeskind-HadasWBK14}.
This relates to the general problem of deciding the precise cost to
assign to evolutionary events in evolutionary models, a recurring
question in the context of parsimony-based approaches in
phylogenetics.

\begin{figure*}
{\centering\includegraphics[width=1.00\linewidth]{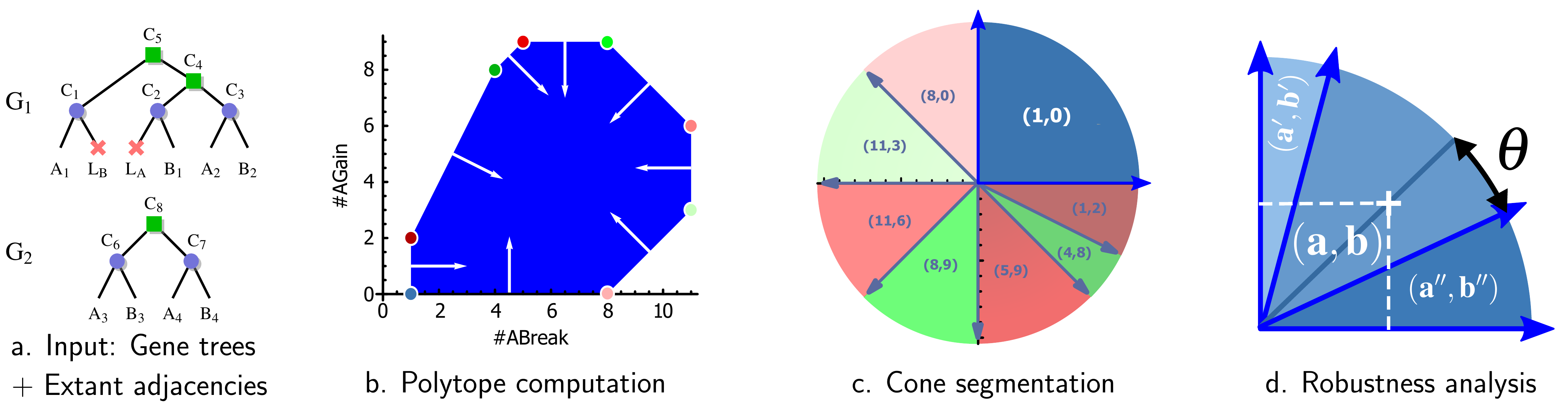}}
\caption{Outline of our method for assessing the robustness of an 
    evolutionary scenario: Starting from
two reconciled gene trees and a set of extant adjacencies (a.), the
polytope of parsimonious signatures is computed (b.). Its normal
vectors define a segmentation of the space of cost schemes into cones
(c.), each associated with a signature. Here, the positive quadrant is
fully covered by a single cone, meaning that the parsimonious
prediction does not depend on the precise cost scheme. In general
(d.), the robustness of a prediction (here, obtained using the $(1,1)$
scheme) to perturbations of the scheme can be measured as the smallest
angle $\theta$ such that a cost scheme at angular distance $\theta$ no
longer predicts the signature \((a,b)\).}\label{fig:method}
\end{figure*}

This motivates the precise questions tackled in this work. First, how robust 
is a parsimonious evolutionary scenario to a change of the costs associated 
to adjacency gains and breaks? Similarly, how robust is an inferred parsimonious 
gene adjacency to a change in these costs? 
We address this problem using a methodology that has been formalized into a 
rigorous algebraic framework by Pachter and
Sturmfels~\citep{PachterS2005,PachterS2004a,PachterS2004b}, that we
refer to as the \emph{polytope approach}. Its main features,
summarized in Fig.~\ref{fig:method} for assessing the robustness of 
evolutionary scenarios, are (1) associating each evolutionary scenario to 
a \emph{signature}, a vector of two integers $(g,b)$ where $g$ is the number 
of adjacency gains and $b$ the number of adjacency breaks; and (2) 
partitioning the space of cost schemes into convex regions such that, for 
all the cost schemes within a region, all optimal solutions obtained with 
such cost schemes have the same signature. This partition can be computed 
by an algorithm that is a direct translation of the DP algorithm into a 
polytope framework. Furthermore, the same framework can be extended to 
assess the robustness of inferred parsimonious ancestral adjacencies.


%% file: preliminaries_isbra.tex


A {\em phylogeny} is a rooted tree which describes the evolutionary
relationships of a set of elements (species, genes, \dots) represented
by its nodes: internal nodes correspond to ancestral elements, leaves
to extant elements, and edges represent direct descents between
parents and children. For a node $v$ of a phylogeny, we denote by
$s(v)$ the species it belongs to. For a tree $T$ and a node $x$ of
$T$, we denote by $T(x)$ the subtree rooted at $x$. If $x$ is an
internal node, we assume it has either one child, denoted by $a(x)$,
or two children, denoted by $a(x)$ and $b(x)$. 

\begin{figure}
  {\centering \includegraphics[scale=0.55]{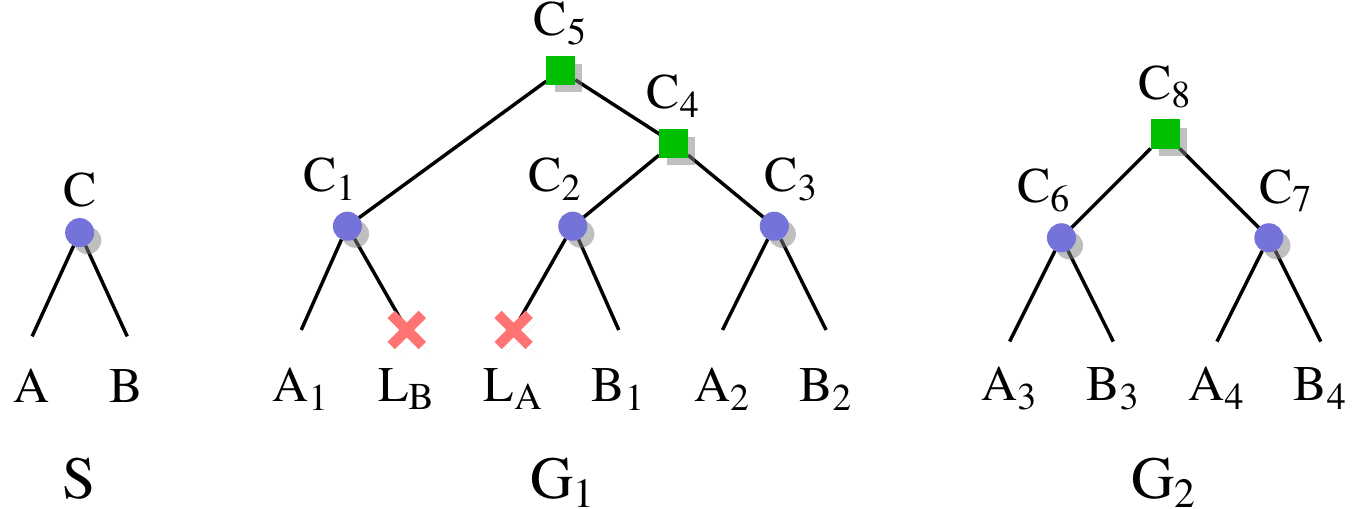}
  \includegraphics[scale=0.55]{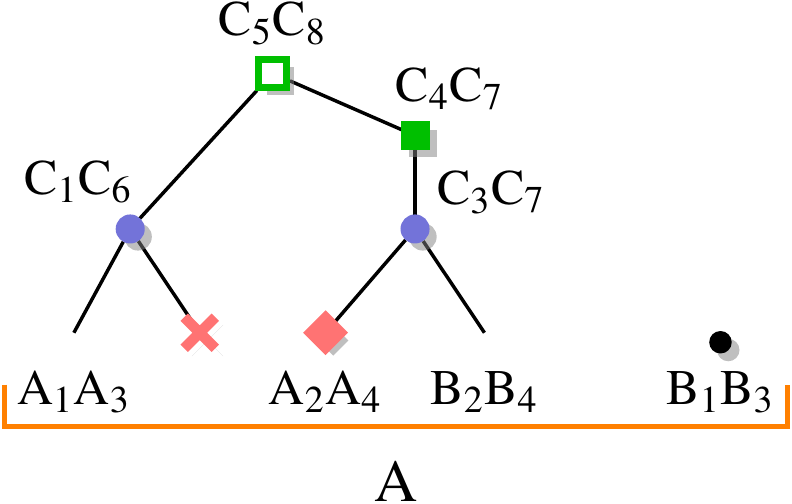}}
  \vspace{-1em}
  \caption{A species tree $S$, with two extant species $A$ and $B$ and
    an ancestral species $C$.  Two reconciled gene trees $G_1$ and
    $G_2$, with four extant genes in genome $A$, four extant genes in
    genome $B$ and three ancestral genes in genome $C$. The set of
    extant gene adjacencies is $(A_1A_3,B_1B_3,B_2B_4)$. An adjacency
    forest $A$ composed of two adjacency trees. Blue dots represent
    speciation nodes. Leaves are extant species/genes/adjacencies,
    except the one labeled by a red cross (gene loss) or a red diamond
    (adjacency breaks). Green squares are (gene or adjacency)
    duplication nodes. Gene labels refer to the species they belong
    to. Every node of the adjacency tree is labeled by a gene adjacency.  Figure adapted
    from~\citep{DBLP:journals/bioinformatics/BerardGBSDT12}.}
  \label{fig:trees}
\end{figure}

\paragraph{Species tree and reconciled gene trees.}
A species tree $S$ is a binary tree that describes the evolution of a
set of species from a common ancestor through the mechanism
of \emph{speciation}.  A reconciled gene tree is a binary tree that
describes the evolution of a set of genes, called a \emph{gene
family}, within a given species tree $S$, through the evolutionary
mechanisms of speciation, \emph{gene duplication} and {\em gene
loss}. Therefore, each leaf of a gene tree $G$ represents either a
gene loss or an an extant gene, while each internal node represents an
ancestral gene.
In a reconciled gene tree, we associate every ancestral gene (an
internal node $g$) to an evolutionary event $e(g)$ that leads to the
creation of the two children $a(g)$ and $b(g)$: $e(g)$ is a speciation
(denoted by $\Spec$) if the species pair $\{s(a(g)),s(b(g))\}$ is
equal to the species pair $\{a(s(g)),b(s(g))\}$, $s(a(g))\neq
s(b(g))$, or a gene duplication ($\GDup$) if
$s(a(g))=s(b(g))=s(g)$. If $g$ is a leaf, then $e(g)$, as stated before, indicates either 
a gene loss ($\GLos$) or an extant gene ($\Extant$), in which case
$e(g)$ is not an evolutionary event \emph{stricto sensu}. A
\emph{pre-speciation} ancestral gene is an internal node $g$ such that
$e(g)=\Spec$. See Fig.~\ref{fig:trees}  for an
illustration.

\paragraph{Adjacency trees and forests.}
We consider now that we are given two reconciled gene trees $G_1$ and
$G_2$, representing two gene families evolving within a species tree
$S$.  A \emph{gene adjacency} is a pair of genes (one from $G_1$ and
one from $G_2$) that appear consecutively along a chromosome, for a
given species, ancestral or extant. Gene adjacencies evolve within a
species tree $S$ through the evolutionary events of speciation, gene
duplication, gene loss (these three events are modeled in the
reconciled gene trees), and
\emph{adjacency duplication} ($\ADup$), \emph{adjacency loss}
($\ALos$) and \emph{adjacency break} ($\ABreak$), that are
adjacency-specific events.

Following the model introduced
in~\citep{DBLP:journals/bioinformatics/BerardGBSDT12}, we represent
such an evolutionary history using an \emph{adjacency forest},
composed of \emph{adjacency trees}. An adjacency tree represents the
evolution of an ancestral gene adjacency (located at the root of the
tree) through the following events: (1)
The duplication of an adjacency $\{g_1,g_2\}$, where $g_1$ and
  $g_2$ are respectively genes from $G_1$ and $G_2$ such that
  $s(g_1)=s(g_2)$, follows from the simultaneous duplication of both
  its genes $g_1$ and $g_2$ (so $e(g_1)=e(g_2)=\GDup$), resulting in
  the creation of two distinct adjacencies each belonging to
  $\{a(g_1),b(g_1)\}\times\{a(g_2),b(g_2)\}$;
(2)The loss of an adjacency, which can occur due to
  several events, such as the loss of exactly one of its genes (gene
  loss, $\GLos$), the loss of both its genes (adjacency loss, $\ALos$)
  or a genome rearrangement that breaks the contiguity between the two
  genes (adjacency break, $\ABreak$);
(3) The creation/gain of an adjacency (denoted by $\AGain$), for
  example due to a genome rearrangement, that results in the creation
  of a new adjacency tree whose root is the newly created adjacency.

With this model, one can model the evolution of two gene families
along a species phylogeny by a triple $(G_1,G_2,A)$: $G_1$ and $G_2$
are reconciled gene trees representing the evolution of these families
in terms of gene-specific events and $A$ is an adjacency forest
consistent with $G_1$ and $G_2$.  Similar to species trees and
reconciled gene trees, internal nodes of an adjacency tree are
associated to ancestral adjacencies, while leaves are associated to
extant adjacencies or lost adjacencies (due to a gene loss, adjacency
loss or adjacency break), and are labeled by evolutionary events. The
label $e(v)$ of an internal node $v$ of an adjacency forest $A$
belongs to $\{\Spec,\GDup,\ADup\}$, while the label $e(v)$ of a leaf
belongs to $\{\Extant,\GLos,\ALos,\ABreak\}$, as shown in 
Fig.~\ref{fig:trees}.

\paragraph{Signatures, descriptors and parsimonious scenarios.}
The \emph{signature} of an adjacency forest $A$ is an ordered pair of
integers $\Sig(A)=(g_A,b_A)$ where $g_A$ (resp. $b_A$) is the number
of adjacency gains (resp. adjacency breaks) in $A$.  A \emph{cost
scheme} is a pair $\mathbf{x}=(x_0,x_1)$ of non-negative real numbers,
where $x_0$ is the cost of an adjacency gain and $x_1$ the cost of an
adjacency break. The \emph{cost} of an adjacency forest $A$ for a
given cost scheme $\mathbf{x}$ is the number $S(A)=x_0\times
g_A+x_1\times b_A$. The adjacency forest $A$ in an evolutionary
scenario $(G_1,G_2,A)$ is
\emph{parsimonious for $\mathbf{x}$} if there is no other evolutionary
scenario $(G_1,G_2,B)$ such that $S(B)<S(A)$. The signature
the adjacency forest $A$ in Fig.~\ref{fig:trees} is $(1,1)$, and this adjacency forest is
parsimonious for the cost scheme $(1,1)$.

A \emph{descriptor} of a scenario is a boolean or integer valued
feature of the solution which does not contribute to the cost of the
scenario, but rather represents a feature of a scenario.  For instance,
the presence/absence of an ancestral adjacency in a given adjacency
forest \(A\) can be described as a boolean. Given \(k\) descriptors
\(a_{1},\ldots,a_{k}\), we define an \emph{extended signature} of
a scenario \(A\) as a tuple
\(\Sig_{a_{1},\ldots,a_{k}}\left(A\right)=\left(g,b,s_{a_{1}},\ldots,s_{a_{k}}\right)\),
where \(g,b\) are the numbers of adjacency gains and breaks in
\(A\) respectively, and \(s_{a_{i}}\) is the value of the descriptor
$a_i$ for \(A\).

\paragraph{The \DeCo{} algorithm.} B\'erard \emph{et al.}
\cite{DBLP:journals/bioinformatics/BerardGBSDT12} showed that, given a
pair of reconciled gene trees $G_1$ and $G_2$, a list $L$ of extant
gene adjacencies, and a cost scheme $\mathbf{x}$, one can use a DP algorithm to compute an evolutionary scenario $(G_1,G_2,A)$, 
where $A$
is a parsimonious adjacency forest such that $L$ is exactly the set of
leaves of $A$ labeled $\Extant$.  The \DeCo{} algorithm computes, for
every pair of nodes $g_1$ (from $G_1$) and $g_2$ (from $G_2$) such
that $s(g_1)=s(g_2)$, two quantities $c_1(g_1,g_2)$ and
$c_0(g_1,g_2)$, that correspond respectively to the cost of a
parsimonious adjacency forest for the pairs of subtrees $G(g_1)$ and
$G(g_2)$, under the hypothesis that $g_1$ and $g_2$ form (for
$c_1$) or do not form (for $c_0$) an ancestral adjacency. As usual in
dynamic programming along a species tree, the cost of a parsimonious
adjacency forest for $G_1$ and $G_2$ is given by
$\min(c_1(r_1,r_2),c_0(r_1,r_2))$ where $r_1$ is the root of $G_1$ and
$r_2$ the root of $G_2$. In~\cite{DBLP:conf/wob/ChauvePZ14}, we recently
generalized \DeCo{} into a DP algorithm \DeClone{} that allows one to
explore the space of all possible adjacency evolutionary scenarios for
a given cost scheme.

\paragraph{Robustness problems.}\label{sec:robustness}
The first problem we are interested in is the \emph{signature
  robustness problem}. A signature $\Sig=(g,b)$ is parsimonious for
a cost scheme $\mathbf{x}$ if there exists at least one adjacency
forest $A$ that is parsimonious for $\mathbf{x}$ and has signature
$\Sig(A)=\Sig$. The robustness of the signature $\Sig$ is defined
as the difference between $\mathbf{x}$ and the closest cost scheme for 
which $\Sig$ is no longer parsimonious. To measure this
difference, we rely on a geometric representation of a
cost scheme. Assuming that a cost scheme
$\mathbf{x}=(x_0,x_1)\in\mathbb{R}^{2}$ provides sufficient
information to evaluate the cost of an adjacency forest, the
predictions under such a model remain unchanged upon multiplying
$\mathbf{x}$ by any positive number, allowing us to assume that
$\norm{\mathbf{x}}=1$ without loss of generality. So 
$\mathbf{x}=(x_0,x_1)$ can be summarized as an angle $\theta$
(expressed in radians), and the difference between two cost schemes is
indicated by their associated angular distance.

However, signatures only provide a quantitative summary of the
evolutionary events described by a parsimonious adjacency forest.  In
particular, signatures discard any information about predicted sets of
ancestral adjacencies.  We address the robustness of inferred 
parsimonious adjacencies through the \emph{parsimonious adjacency 
robustness problem}. Let
$a=(g_{1},g_{2})$ be an ancestral adjacency featured in a parsimonious
adjacency forest for a cost scheme $\mathbf{x}$. We say that $a$ is
parsimonious for a cost scheme $\mathbf{y}$ if $a$ belongs to every
adjacency forest that is parsimonious for $\mathbf{y}$. The robustness
of $a$ is defined as the angular distance from $\mathbf{x}$
to the closest cost scheme $\mathbf{y}$ for which $a$ is no longer
parsimonious. 


%% file: methods_isbra.tex

If the signature for a given adjacency forest \(A\) is given by the
vector \(\Sig\left(A\right)=\left(g,b\right)\), and the cost scheme is
given by the vector \(\mathbf{x}=\left(x_{0},x_{1}\right)\), then the
parsimony cost of \DeCo{} can be written as the inner product
\(\langle\mathbf{x},\Sig\left(A\right)\rangle =  g\times x_0 + b\times x_1\).  
\DeCo{} computes the following quantity for a pair of gene trees \(G_{1}\) and \(G_{2}\).

\vspace*{-2mm}
\begin{equation}
    c(G_1,G_2) = \min_{A\in\mathcal{F}(G_1,G_2)}\langle\mathbf{x},\Sig\left(A\right)\rangle, \label{eq:objective}
\end{equation}

\vspace*{-2mm}\noindent
where \(\mathcal{F}(G_1,G_2)\) denotes the set of all possible adjacency
forests that can be constructed from \(G_{1}\) and \(G_{2}\),
irrespective of the cost scheme.

For a given adjacency forest $A$, we will consider a single
descriptor \(a\), indicating the presence or absence of an ancestral
adjacency \(a=\left(g_{1},g_{2}\right)\in G_{1}\times G_{2}\) in $A$,
where \(s_{a}=1\) if it is present in \(A\), and \(0\) otherwise.
Since, by definition, a descriptor does not contribute to the cost,
when considering the robustness of specific adjacencies, we will
consider cost schemes of the
form \(\mathbf{x}=\left(x_{0},x_{1},0\right)\), and \DeCo{} will
compute Eq. \eqref{eq:objective} as usual. 

For a given cost scheme \(\mathbf{x}\), two adjacency
forests \(A_{1}\) and \(A_{2}\) such that $\Sig(A_1)=\Sig(A_2)$ will have
the same associated cost. We can thus define an equivalence class
in \(\mathcal{F}(G_1,G_2)\) based on the signatures. However, for a
given potential ancestral adjacency \(a=\left(g_{1},g_{2}\right)\in
G_{1}\times G_{2}\), the adjacency forests in this equivalence class
may have different extended signatures, differing only in the last
coordinate. Thus, there may be two adjacency forests \(A_{1}\) and
\(A_{2}\) with extended signatures \(\left(g,b,1\right)\) and
\(\left(g,b,0\right)\) respectively, and they will have the same
cost for all cost schemes. Evolutionary scenarios with the same
extended signature also naturally form an equivalence class in
\(\mathcal{F}(G_1,G_2)\).

\paragraph{Convex polytopes from signatures.}
Let us denote the set of signatures of all scenarios in
\(\mathcal{F}(G_1,G_2)\) by \(\Sig\left(\mathcal{F}(G_1,G_2)\right)\),
and the set of extended signatures for a given adjacency \(a\) by
\(\Sig_{a}\left(\mathcal{F}(G_1,G_2)\right)\). Each of these is a
point in \(\mathbb{R}^{d}\), where \(d=2\) for signatures and \(d=3\)
for extended signatures. In order to explore the parameter space of
parsimonious solutions to \DeCo{}, we use these sets of points to
construct a \emph{convex polytope} in \(\mathbb{R}^{d}\).  A convex
polytope is simply the set of all convex combinations of points in a
given set, in this case the set of signatures or extended
signatures~\citep{PachterS2005}. Thus, for each pair of gene trees
\(G_{1},G_{2}\) and a list of extant adjacencies, we can theoretically
construct a convex polytope in \(\mathbb{R}^{2}\) by taking the convex
combinations of all signatures in
\(\Sig\left(\mathcal{F}(G_1,G_2)\right)\). This definition generalizes
to a convex polytope in \(\mathbb{R}^{3}\) when extended signatures
\(\Sig_{a}\left(\mathcal{F}(G_1,G_2)\right)\) are considered for some
ancestral adjacency $a$.  Viewing the set of evolutionary scenarios as
a polytope allows us to deduce some useful properties:
\begin{enumerate}
\item Any (resp. extended) signature that is parsimonious for some
  cost scheme \(\mathbf{x}\) lies on the surface of the polytope;
\item If a (resp. extended) signature is parsimonious for two cost
  schemes \(\mathbf{x}\) and \(\mathbf{x}'\), then it is also
  parsimonious for any cost scheme {\em in between} (i.e. for any
  convex combination of \(\mathbf{x}\) and
  \(\mathbf{x}'\)).\label{prop:cone}
\end{enumerate}
Traditionally, a polytope is represented as a set of inequations,
which is inappropriate for our intended application. Therefore, we
adopt a slighty modified representation, and denote the polytope of
\(\mathcal{F}\left(G_{1},G_{2}\right)\) as the list of signatures that
are represented within \(\mathcal{F}\left(G_{1},G_{2}\right)\) and lie
on the convex hull of the polytope.

A \emph{vertex} in a polytope is a signature (resp. extended
signature) which is parsimonious for some cost scheme. The domain of
parsimony of a vertex \(\mathbf{v}\) is the set of cost schemes for
which \(\mathbf{v}\) is parsimonious.  From Property~\ref{prop:cone},
the domain of parsimony for a vertex \(\mathbf{v}\) is a \emph{cone}
in \(\mathbb{R}^d\), formally defined as:
\begin{align}
    Cone\left(\mathbf{v}\right)&=\left\{\mathbf{x}\in\mathbb{R}^{d}:\langle \mathbf{x},\mathbf{v}\rangle\leq \langle
\mathbf{x},\mathbf{w}\rangle\; \forall\ \mathbf{w}\in P\right\}.
\end{align}
The set of cones associated with the vertices of a polytope form a
partition of the cost schemes space~\citep{PachterS2005}, which allows
us to assess the effect of perturbing the cost scheme on the optimal
solution of \DeCo{} for this cost scheme.

\paragraph{Computing the polytope.}
Building on earlier work on parametric sequence alignment~\citep{DBLP:journals/algorithmica/GusfieldBN94},
Pachter and Sturmfels~\citep{PachterS2004a,PachterS2005} described the
concept of \emph{polytope propagation}, based on the observation that
the polytope of a DP (minimization) scheme can be computed through an
algebraic substitution. Accordingly, any point that lies strictly
within the polytope is suboptimal for any cost scheme, and can be
safely discarded by a procedure that repeatedly computes the {\em
  convex hull} $H(P)$ of the (intermediates) polytopes produced by the
modified DP scheme. In the context of the \DeCo{} DP scheme, the
precise modifications are:
\begin{enumerate}
\item\vspace*{-2mm} Any occurrence of the \(+\) operator is replaced by \(\oplus\),
  the (convex) {\em Minkowski sum} operator, defined for $P_1, P_2$
  two polytopes as \vspace*{-2mm}
    \[ P_1 \oplus P_2 = H(\{p_1 + p_2\mid (p_1,p_2)\in P_1\times P_2\});\]
\item\vspace*{-2mm} Any occurrence of the \(\min\) operator is replaced by \(\Cup\),
  the {\em convex union} operator, defined for $P_1, P_2$ two
  polytopes as \vspace*{-2mm}\[ P_1 \Cup P_2 = H(P_1 \cup P_2);\]
\item\vspace*{-2mm}  Any occurrence of an {\em adjacency gain} cost is replaced by
  the vector \(\left(1,0\right)\) (resp. \(\left(1,0,0\right)\) for
  extended signatures);
\item Any occurrence of an {\em adjacency break} cost is replaced by
  the vector \(\left(0,1\right)\) (resp. \(\left(0,1,0\right)\) for
  extended signatures);
\item (Extended signatures only) An event that corresponds to
  the prediction of a fixed ancestral adjacency \(a\) in a scenario is
  replaced by the vector \(\left(0,0,1\right)\);
\end{enumerate}
By making this substitution, we can efficiently compute the polytope
associated with two input gene trees \(G_{1}\) and \(G_{2}\), having
sizes \(n_1 \) and \(n_2\) respectively, through \(O\left(n_1\times
n_2\right)\) executions of the convex hull procedure. In place of the
integers used by the original minimization approach, intermediate
convex polytopes are now processed by individual operations, and
stored in the DP tables, so the overall time and space complexities of
the algorithm critically depend on the size of the polytopes, i.e. its
number of vertices.  Pachter and Sturmfels proved that, in general,
the number of vertices on the surface of the polytope is
\(O\left(n^{d-1}\right)\), where \(d\) is the number of dimensions,
and \(n\) is the size of the DP table. In our case, the number of
vertices in the 2D polytope associated with simple signatures is in
\(O\left(n_1\times n_2\right)\). This upper bound also holds for
extended signatures, as the third coordinate is a boolean, and the
resulting 3D polytope is in fact the union of two 2D polytopes.  The
total cost of computing the polytope is therefore bounded by
\(O\left(n_1^2\times n_2^2\times \log(n_1\times n_2)\right)\),
e.g. using Chan's convex hull algorithm~\citep{Chan1996}.
As for the computation of the cones, let us note that the cone of a
vertex \(v\) in a given polytope \(P\) is fully delimited by a set of
vectors, which can be computed from \(P\) as the normal vectors,
pointing towards the center of mass of \(P\), of each of the facets in
which \(v\) appears. This computation can be performed as a
postprocessing using simple linear algebra, and its complexity will
remain largely dominated by that of the DP-fuelled polytope
computation.

\paragraph{Assessing signature and adjacency robustness.} 
The cones associated with the polytope of a given instance cover all
the real-valued cost schemes, including those associating negative
costs to events. These later cost schemes are not valid, and so, we
only consider cones which contain at least one positive cost scheme.
Given a fixed cost scheme \(\mathbf{y}\), the vertex associated to the
cone containing this cost scheme corresponds to the signature of all
parsimonious scenarios for this cost scheme. In order to assess the
robustness of this signature, we can calculate the smallest angular
perturbation needed to move from \(\mathbf{y}\) to a cost scheme whose
parsimonious scenarios do not have this signature. This is simply the
angular distance from \(\mathbf{y}\) to the nearest boundary of the
cone which contains it.  Using this methods, we assign a numerical
value to the robustness of the signatures of parsimonious scenarios on
a number of instances for a particular cost scheme.  

In the case of extended signatures
\(\Sig_{a}\left(\mathcal{F}(G_1,G_2)\right)\) for an adjacency \(a\),
the polytope is \(3\)-dimensional. The cones associated with the
vertices, as defined algebraically, now partition $\mathbb{R}^{3}$,
the set of cost schemes \(\left(x_{0},x_{1},x_{2}\right)\), where
\(x_{2}\) indicates the cost of a distinguished adjacency.  Since the
third coordinate is a descriptor, it does not contribute to the cost
scheme, and we therefore restrict our analysis to the
$\mathbb{R}^+\times \mathbb{R}^+\times\{0\}$ subset of the cost scheme
space.  Precisely, we take the intersection of the plane \(x_{2}=0\)
with each cone associated to a vertex \(\left(g,b,s_{a}\right)\), and
obtain the region in which the extended signature
\(\left(g,b,s_{a}\right)\) is parsimonious. This region is a
2D cone.

However, the cost of an extended signature is independent of the
entry in its last coordinate, and there may exist two different
extended signatures \(\left(g,b,0\right)\) and
\(\left(g,b,1\right)\), both parsimonious for all the cost schemes
found in the 2D cone. It is also possible for adjacent 
cones to have different signatures, yet feature a given 
adjacency. The robustness of a given adjacency \(a\) is 
computed from the cones using a greedy algorithm which, 
starting from the cone containing $\mathbf{x}$, explores the adjacent cones in both directions (clockwise/counter-clockwise) 
until it finds one 
that no longer predicts $a$, i.e. is associated with at least one 
signature \(\left(g',b',0\right)\).

%% file: results_isbra.tex


We considered $5,039$ reconciled gene trees and $50,389$ extant gene
adjacencies, forming $6,074$ \DeCo{} instances, with genes taken from
$36$ extant mammalian genomes from the Ensembl database in
2012. In~\citep{DBLP:journals/bioinformatics/BerardGBSDT12}, this 
data was analyzed with \DeCo{}, using the cost scheme $(1,1)$, that
computed a single parsimonious adjacency forest per instance. These
adjacency forests defined $96,482$ ancestral adjacencies (adjacencies
between two pre-speciation genes from the same ancestral species),
covering $112,188$ ancestral genes.

\begin{figure}[t] 
  \includegraphics[width=0.435\textwidth]{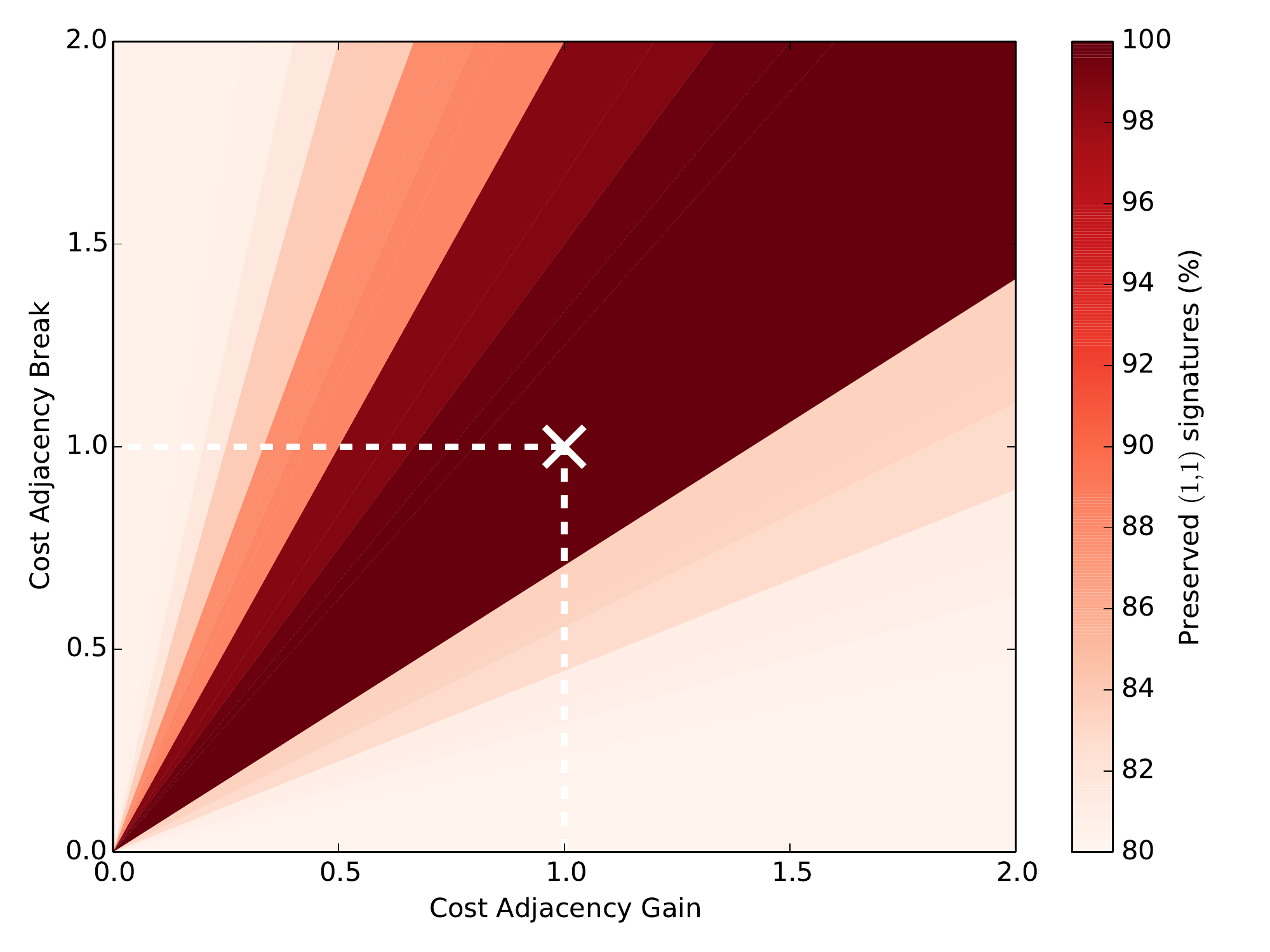}     
  \includegraphics[width=0.515\textwidth]{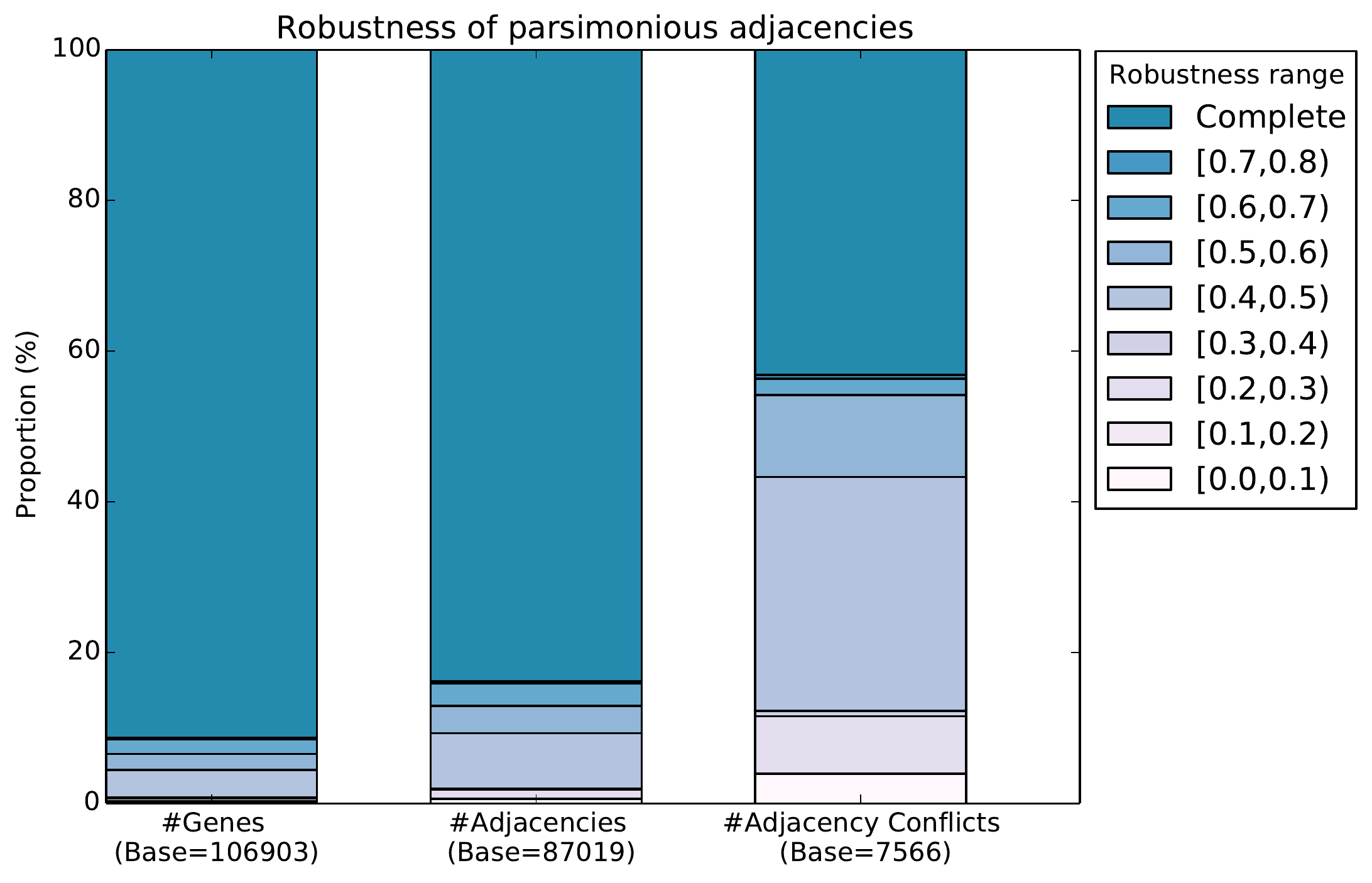}
  \caption{(A) Average robustness of signatures predicted using the
    $(1,1)$ cost scheme.  At each point $(x,y)$, the colour
    indicates the proportion of signatures that are parsimonious,
    and therefore predicted, for the $(1,1)$ cost scheme, and remain
    parsimonious for the $(x,y)$ cost scheme.  (B) Universally
    parsimonious adjacencies and syntenic conflicts. (Left)
    Percentage of ancestral genes present in universally
    parsimonious adjacencies per level of minimum robustness of the
    adjacencies, expressed in radians. (Center) Percentage of
    universally parsimonious adjacencies per level of minimum
    robustness. (Right) Percentage of conserved conflicting
    adjacencies per level of minimum robustness.  }
  \label{fig:2D}
  \label{fig:rob_adj}
\end{figure}

We first considered all $6,074$ instances, and computed for each
signature the robustness of the parsimonious signature obtained with
the cost scheme $(1,1)$. Interestingly, we observe
(Fig.~\ref{fig:2D}(A)) that for more than half of the instances, the
parsimonious signature is robust to a change of cost scheme, as the
associated cone is the complete first quadrant of the real plane. On
the other hand, for $945$ instances the parsimonious signature for the
cost scheme $(1,1)$ is not robust to any change in the cost scheme;
these cases correspond to interesting instances where the cost scheme
$(1,1)$ lies at the border of two cones, meaning that two parsimonious
signatures exist for the cost scheme $(1,1)$, and any small change of
cost scheme tips the balance towards one of these two signatures. More
generally, as revealed by Fig.~\ref{fig:2D}(A), we observe an extreme
robustness of parsimonious signatures: there is a \(\sim80\%\) overlap
between the sets of signatures that are parsimonious for any
(positive) cost scheme, and for the \((1,1)\) cost scheme. 
This observation supports the notion of a sparsely-populated search space 
for attainable signatures. In this vision, signatures are generally isolated,
making it difficult to trade adjacency gains for breaks (or vice-versa) in order 
to challenge the \((1,1)\)-parsimonious prediction. We hypothesize that such 
a phenomenon is essentially combinatorial, as extra 
adjacency gains typically lead, through duplications to more subsequent adjacency 
breaks.


Next, to evaluate the stability of the total number of evolutionary events
inferred by parsimonious adjacency forests, we recorded two counts of
evolutionary events for each instance: the number of syntenic events
(adjacencies gains and breaks) of the parsimonious signature (called
the \emph{parsimonious syntenic events count}), and the maximum number
of syntenic events taken over all signatures that are parsimonious for
some cost scheme (called the \emph{maximum syntenic events count}). We
observe that the average parsimonious (resp. maximum) syntenic events
count is $1.25$ (resp. $1.66$). This shows a strong robustness of the
(low) number of syntenic events to changes in the cost scheme.

We then considered the robustness of individual ancestral
adjacencies. Using the variant \DeClone{} of \DeCo{} that explores the
set of all evolutionary scenarios~\cite{DBLP:conf/wob/ChauvePZ14}, we
extracted, for each instance, the set of ancestral adjacencies that
belong to all parsimonious solutions for the cost scheme $(1,1)$, and
computed their robustness as defined in the previous sections. This
set of ancestral adjacencies contains $87,019$ adjacencies covering
$106,903$ ancestral genes. The robustness of
these adjacencies is summarized in Fig.~\ref{fig:rob_adj}(B, left and
center columns).  It is interesting to observe that few adjacencies
have a low robustness, while, conversely, a large majority of the
universally parsimonious adjacencies are completely robust to a change
of cost scheme ($97,593$ out of $106,639$). This suggests that
the \DeCo{} model of parsimonious adjacency forests is robust, and
infers highly supported ancestral adjacencies, which is reasonable
given the relative sparsity of genome rearrangements in evolution
compared to smaller scale evolutionary events.

Besides the notions of robustness, an indirect validation criterion
used to assess the quality of an adjacency forest is the limited
presence of \emph{syntenic conflicts}. An ancestral gene is said to
participate in a syntenic conflict if it belongs to three or more
ancestral adjacencies, as a gene can only be adjacent to at most two
neighboring genes along a chromosome. An ancestral adjacency
participates in a syntenic conflict if it contains a gene that does.
Among the ancestral adjacencies inferred by \DeCo{}, $16,039$
participate in syntenic conflicts, covering $5,817$ ancestral
genes. This represents a significant level of syntenic conflict and a
significant issue in using \DeCo{} to reconstruct
ancestral gene orders. It was observed that selecting  
universally parsimonious ancestral adjacencies, as done in the previous analysis, 
significantly reduced the number of syntenic conflicts, as almost all discarded ancestral
adjacencies participated in syntenic conflicts.  Considering syntenic conflicts, we observe
(Fig.~\ref{fig:rob_adj}(B, right column) a positive result, i.e. that
filtering by robustness results in a significant decrease of the ratio
of conflicting adjacencies. However, even with robust universally
parsimonious ancestral adjacencies, one can observe a significant
number of adjacencies participating in syntenic conflicts. We discuss
these observations in the next section.

%% file: discussion_isbra.tex
From an application point of view, the ability to exhaustively explore
the parameter space leads to the observation that, on the considered
instances, the \DeCo{} model is extremely robust. Even taking
parsimonious signatures that maximize the number of evolutionary
syntenic events (i.e. considering cost schemes that lead to the
maximum number of events) results in an average increase of roughly
$33\%$ events ($1.25$ to $1.66$), and stays very low, much lower than
gene specific events such as gene duplications (average of $3.38$
event per reconciled gene tree). This is consistent with the fact that
for rare evolutionary events such as genome rearrangements, a
parsimony approach is relevant, especially when it can be complemented
by efficient algorithms to explore slightly sub-optimal solutions,
such as \DeClone{}, and to explore the parameter space. In terms of
direct applications of the method developed here and
in~\citep{DBLP:conf/wob/ChauvePZ14}, gene-tree based reconstruction of
ancestral gene orders comes to
mind~\cite{DBLP:books/daglib/p/ChauveEGST13}; more precisely,
ancestral adjacencies could be determined and scored using a mixture of
their Boltzmann probability (that can be computed efficiently using
\DeClone{}) and robustness to changes of the cost scheme, and
conflicts could be cleared out independently and efficiently for each
ancestral species using the algorithm
of~\cite{DBLP:journals/bmcbi/ManuchPWCT12} for example.

An interesting observation is that even the set of ancestral
adjacencies that are universally-parsimonious and robust to changes in
the cost scheme contains a significant number of adjacencies
participating in syntenic conflict. We conjecture that the main reason
for syntenic conflicts is in the presence of a significant number of
erroneous reconciled gene trees. This is supported by the observation
that the ancestral species with the highest number of syntenic
conflict are also species for which the reconciliation with the
mammalian species tree resulted in a significantly larger number of
genes than expected (data not shown). This points clearly to errors in
either gene tree reconstruction or in the reconciliation with the
mammalian species phylogeny, which tends to assign wrong gene
duplications in some specific species, resulting an inflation of the
number of genes, especially toward the more ancient
species~\citep{Hahn-bias2007}. It would be interesting to see if the
information about highly suported conflicting adjacencies can be used
in reconciled gene tree correction.

From a methodological point of view, we considered here extended
signatures for a single ancestral adjacency at a time. It would be
natural to extend this concept to the more general case of several
ancestral adjacencies considered at once. We conjecture that this case
can be addressed without an increase in the asymptotic complexity of
computing the polytope; this problem will be considered in the full
version of the present work.  Next, there exists another way to
explore the parameter space of a dynamic programming phylogenetic
algorithm.  It consists of computing the \emph{Pareto-front} of the input
instance~\citep{DBLP:journals/bioinformatics/Libeskind-HadasWBK14,DBLP:conf/eccb/SauleG14}, rather than optimal signatures
for classes of cost schemes. A signature \(v\) is said to be \emph{Pareto-optimal} if there is no other signature whose entries
are equal or smaller than the
corresponding entries in \(v\), and is strictly smaller at at least
one coordinate. The Pareto-front is the set of all Pareto-optimal
signatures, and can be efficiently computed by dynamic
programming~\citep{DBLP:journals/bioinformatics/SchnattingerSK13,DBLP:conf/eccb/SauleG14,DBLP:journals/bioinformatics/Libeskind-HadasWBK14}. The
Pareto-front differs from the approach we describe in the present work
in several aspects.  An advantage of the Pareto-front is that it is a
notion irrespective of the type of cost function being used. This
contrasts with the polytope propagation technique, which requires that
the cost function be a linear combination of its terms. However, so
far, the Pareto-approach has only been used to define a partition of
the parameter space when the cost function is restricted to be
linear/affine, and it remains to investigate the difference with the
polytope approach in this case. 